\documentclass[Journal]{IEEEtran}
\usepackage{color}
\usepackage{graphicx}
\usepackage{amsmath}
\usepackage{mathtools}
\usepackage{cite}
\usepackage{enumitem}
\usepackage{textcomp}

\ifCLASSINFOpdf
\else
\fi

\renewcommand{\baselinestretch}{0.95}
\begin{document}
%
\title{Parallel FPGA Router using Sub-Gradient method and Steiner tree}
%
%
%

\author{Rohit~Agrawal, Chin~Hau~Hoo, Kapil~Ahuja, and Akash~Kumar
\thanks{R. Agrawal and K. Ahuja are with Discipline of Computer Science and Engineering, Indian Institute of Technology Indore, India e-mail: kahuja@iiti.ac.in.}
\thanks{C. H. Hoo is with Department of Electrical and Computer Engineering, National University of Singapore, Singapore.}
\thanks{A. Kumar is with Centre for Advancing Electronics, Technische Universit\"at Dresden, Germany}}
\maketitle

\begin{abstract}
In the FPGA (Field Programmable Gate Arrays) design flow, one of the most time-consuming step is the routing of nets. Therefore, there is a need to accelerate it. In \cite{Hau}, the authors have developed a Linear Programming (LP) based framework that parallelizes this routing process to achieve significant speedups (the resulting algorithm is termed as ParaLaR). However, this approach has certain weaknesses. Namely, the constraints violation by the solution and a local minima that could be improved. We address these two issues here. 

In this paper, we use the LP framework of \cite{Hau} and solve it using the Primal-Dual sub-gradient method that better exploits the problem properties. We also propose a better way to update the size of the step taken by this iterative algorithm. We perform experiments on a set of standard benchmarks, where we show that our algorithm outperforms the standard existing algorithms (VPR \cite{Betz} and ParaLaR). 

We achieve upto 22\% improvement in the constraints violation and the standard metric of the minimum channel width when compared with ParaLaR (which is same as in VPR). We achieve about 20\% savings in another standard metric of the total wire length (when compared with VPR), which is the same as for ParaLaR. Hence, our algorithm achieves minimum value for all the three parameters. Also, the critical path delay for our algorithm is almost same as compared to VPR and ParaLaR. On an average, we achieve relative speedups of $3$ times when we run a parallel version of our algorithm using $4$ threads.
\end{abstract}


%
\IEEEpeerreviewmaketitle

\vspace{-3mm}
\section{Introduction}
%
%
%
%
\IEEEPARstart{A}{}ccording to the Moore's law, the number of transistors in an integrated circuit is doubling approximately every two years. In the FPGA design flow, the routing of nets (which are a collection of two or more interconnected components) is one of the most time consuming step. Hence, there is a need to develop fast routing algorithms that tackle the problem of the increasing numbers of transistors per chip, and subsequently, the increased runtime of FPGA CAD tools. This can be achieved in two ways. First, by parallelizing the routing algorithms for hardware having multiple cores. However, the pathfinder algorithm \cite{Murchi}, which is one of the most commonly used FPGA routing algorithm is intrinsically sequential. Hence, this approach seems inappropriate for parallelizing all types of FPGA routing algorithms. 

Second, instead of compiling the entire design together, the users can partition their design, compile partitions progressively, and then assemble all the partitions to form the entire design. Some existing works have proposed this approach \cite{Cabral, Gort}. However, the routing resources required by one partition may be held by another partition, i.e. there is no guarantee to have balanced partitions. In other words, in this approach, there is a need to tackle the difficulties arising in sharing of routing resources. 

The authors in \cite{Hau} overcome the limitations of existing approaches by formulating the FPGA routing problem as an optimization problem. Here, the objective function is linear and the decision variables can only have binary values. Hence, the FPGA routing problem is converted to a LP minimization problem (LP is an optimization technique in which the objective function and the constraints are linear). In this LP, the dependencies that prevent the nets from being routed in parallel are examined and relaxed by using Lagrange relaxation multipliers. The relaxed LP is solved in a parallel manner by the sub-gradient method and the Steiner tree algorithm, which is called ParaLaR. 

This parallelization gives significant speedups. However, in this approach, the sub-gradient method is used in a standard way that does not always gives feasible solution (i.e. some constraints are violated). Further, by this approach, although the metric of total wire length is reduced as compared to VPR \cite{Betz} (which is another commonly used algorithm for FPGA routing), but the metric of minimum channel width needs to be further improved. 

There are many variants of the sub-gradient method and a problem specific method gives better result. In this paper, we use the same framework as for ParaLaR, but use an adapted sub-gradient method. Our approach substantially solves the above two problems. That is, as compared to results in \cite{Hau}, the number of infeasible solutions and the minimum channel width requirement both reduce by about 22\% (which is same as in VPR). As compared to VPR, we save about 20\% in the total wire length as well, which is the same improvement as obtained in ParaLaR. Hence, our algorithm achieves minimum value for all the three parameters. The critical path delay for our algorithm is almost same as compared to VPR and ParaLaR. On an average, we achieve relative speedups of $3$ times when we run a parallel version of our algorithm using $4$ threads.

The rest of this paper is organized as follows: Section \ref{Sec:Formulation of optimization problem} describes the formulation of the FPGA routing as an optimization problem. Section \ref{Sec:Implementation} explains the  implementation of our proposed approach. Section \ref{Sec:Experimental Results} presents experimental results. Finally, Section \ref{Sec:Conclusion} gives conclusions and discusses future work.


 




\section{Formulation of the optimization problem}\label{Sec:Formulation of optimization problem}
The routing problem in FPGA or electronic circuit design is a standard problem that is formulated as a weighted grid graph $G(V,E)$ of certain set of vertices \textit{V}  and edges \textit{E}, where a cost is associated with each edge. In this grid graph, there are three types of vertices; the net vertices, the Steiner vertices, and the other vertices. A net is represented as a set $N \subseteq V$, and set N consist of all net vertices. A Steiner vertex is not part of the net vertices but it is used to construct the net tree, which is the route of a net (i.e. a sub-tree \textit{T} of the graph \textit{G}). A net tree is also called a Steiner tree.

\begin{figure}[h!]
	\centering
	\includegraphics[width=0.8\linewidth]{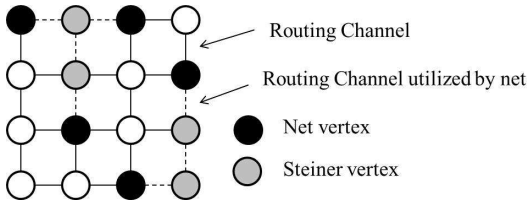}
	\caption{$A \ 4\times4 \ grid \: graph$.}
	\label{fig:grid graph}
\end{figure}

Fig. \ref{fig:grid graph} shows an example of $4\times4$ grid graph. In this figure, the black color circles represent the net vertices; the gray color circles represent the Steiner vertices; and the white color circles are the other vertices. The horizontal and the vertical lines represent the edges (as above, these edges have a cost associated with them but that is not marked here). Two net trees are shown by dotted edges.

The number of nets and the set of vertices belonging to each net is given. The objective here is to find a route for each net such that the union of all the routes will minimize the total path cost of the graph \textit{G}. The goal here is to also minimize the channel width requirement of each edge. Both these objectives are explained in detail below, after (\ref{eq:LP}).

To achieve the above two objectives, the problem of routing of nets is formulated as a LP problem given as follows \cite{Hau}:

\begin{align}\label{eq:LP} 
& \underset{x_{e,i}}{Minimize} \sum_{i=1}^{N_{nets}}\sum_{e\in E}w_{e}x_{e,i},\\ \notag
& Subject \hspace{0.1cm} to \sum_{i=1}^{N_{nets}}x_{e,i}\le W, \forall e\in E, \\ \notag
& \qquad\qquad\quad A_{i}x_{i}=b_{i}, i=1,2,...N_{nets} \\ \notag
&\qquad\qquad\quad x_{e,i}=0 \hspace{0.1cm}or \hspace{0.1cm} 1.
\end{align}

This optimization problem minimizes the total path cost of FPGA routing, where $N_{nets}$ is the number of nets, \textit{E} is the set of edges, $w_e$ is the cost/ time delay associated with the edge \textit{e}, $x_{e,i}$ is the decision variable that indicates whether an edge (routing channel) \textit{e} is utilized by the net \textit{i} (value 1) or not (value 0), $x_{i}$ is the vector of all $x_{e,i}$ for net $i$ that represents the $i^{th}$ net\textquotesingle s route tree, $A_{i}$ is the node-arch incidence matrix, and $b_{i}$ is the demand/ supply vector. The inequality constraints are the channel width constraints that restrict the number of nets utilizing an edge to a constant \textit{W} (which is iteratively reduced as well; discussed later). The equality constraints guarantee that a valid route tree is formed for each net, and these are implicitly satisfied by our solution approach. 

To find a feasible route for each net efficiently, the above LP should be parallelized. There are two main challenges here, which are discussed next. 

{\vspace{-4mm}
\subsection{The channel width constraints}\label{sec:The channel width constraints}

The first challenge to parallelize the LP given in (\ref{eq:LP}) is created by the channel width constraints. These constraints introduce dependency in the parallelizing process, and therefore, should be eliminated or relaxed (see \cite{Hau} for further details). The Lagrange relaxation \cite{Fisher} is a technique well suited for problems where the constraints can be relaxed by incorporating them into the objective function\footnote{If this is not possible, then a Lagrange heuristic can be developed \cite{Lagrangian heuristics 1}, \cite{Lagrangian heuristics 2}, \cite{Lagrangian heuristics 3}.}. For our problem, $\lambda_{e}$ times the corresponding channel width constraints are added to the original objective function to obtain the modified LP. That is, we have the following \cite{Hau}:
\begin{align}\label{eq:relaxed LP}
& \underset{x_{e,i}, \lambda_{e}}{Minimize} \Bigg(\sum_{i=1}^{N_{nets}}\sum_{e\in E}w_{e}x_{e,i}+\sum_{e\in E}\lambda _{e} \bigg( \sum_{i=1}^{N_{nets}}x_{e,i}-W \bigg)\Bigg),\\ \notag
& Subject \hspace{0.1cm} to \hspace{0.1cm} A_{i}x_{i}=b_{i}, \hspace{0.1cm} i=1,2...N_{nets}. \\ \notag
& \qquad\qquad\quad x_{e,i}=0 \hspace{0.1cm}or \hspace{0.1cm} 1 \quad and \\ \notag
& \qquad\qquad\quad \lambda_{e}\geq0.
\end{align}
This LP is independent of the channel width constraints, and hence, it can be solved in a parallel manner. After rearranging the objective function in (\ref{eq:relaxed LP}), the above modified LP is given as \cite{Hau} 
\begin{align}\label{eq:relaxed LP modified}
& \underset{x_{e,i}, \lambda_{e}}{Minimize} \Bigg(\sum_{i=1}^{N_{nets}}\sum_{e\in E}\left ( w_{e}+\lambda _{e} \right )x_{e,i}-W\sum_{e\in E}\lambda _{e}\Bigg),\\ \notag
& Subject \hspace{0.1cm} to \hspace{0.1cm} A_{i}x_{i}=b_{i}, \hspace{0.1cm} i=1,2...N_{nets}. \\ \notag
& \qquad\qquad\quad x_{e,i}=0 \hspace{0.1cm}or \hspace{0.1cm} 1 \quad and \\ \notag
& \qquad\qquad\quad \lambda_{e}\geq.
\end{align}

In (\ref{eq:relaxed LP modified}), $\left(w_{e}+\lambda_{e}\right)$ is the new cost associated with the edge \textit{e}. 
\vspace{-4mm}
\subsection{The choice of decision variable}\label{sec:The Choice of decision variable}

The second challenge to solve the LP given by (\ref{eq:LP}) or the modified LP given by (\ref{eq:relaxed LP modified}) is created by the decision variables $x_{e,i}$. These decision variables are restricted to take value either $0$ or $1$ (as earlier, if the edge \textit{e} is utilized by the net \textit{i}, then $x_{e,i}=1$ else $x_{e,i}=0$). Thus, this is a binary integer linear program (BILP), which is non-differentiable, and hence, cannot be solved by conventional methods such as the Simplex method \cite{Bartels}, the interior point method \cite{Lustig}, etc. Some methods to solve non-differentiable optimization problems include sub-gradient based methods \cite{Duchi}, the approximation method \cite{Bertsekas}, etc. 


The sub-gradient based methods are commonly used algorithms to minimize non-differentiable convex functions $f(x)$. These are iterative in nature that update the variable \textit{x} as $x^{k+1}=x^{k}-\alpha^{k}g^{k}$, where $\alpha^k$ and $g^k$ are the step size and a sub-gradient of the objective function, respectively, at iteration $k$. In \cite{Hau}, the LP given in (\ref{eq:relaxed LP modified}) is not solved directly by a sub-gadient based method but only the Lagrange relaxation multipliers are obtained by it. 
After this (i.e. after solving Lagrange relaxation multipliers), the minimum Steiner tree algorithm is used in a parallel manner for FPGA routing. Here, the decision variables $x_{e,i} \ \forall i\in {1, 2, ..., N_{nets}}$ can have only binary values. Just using a sub-gradient method will not always give binary solutions. Moreover, using a Steiner tree algorithm helps us in achieving feasible routing (equality constraints are implicitly satisfied).

{\vspace*{-3mm}
\section{Implementation}\label{Sec:Implementation}
There are many variants of sub-gradient based methods such as the Projected sub-gradient method \cite{Duchi}, the Primal-Dual sub-gradient method \cite{Boyd}, the Conditional sub-gradient method \cite{Guta}, the Deflected sub-gradient method \cite{Guta}, etc. In \cite{Hau}, authors use the Projected sub-gradient method, where the Lagrange relaxation multipliers are calculated as $\lambda^{k+1}=\max\left(0,\lambda^{k}+\alpha^{k}h\right)$. Here, $\lambda^{k}$ and $\lambda^{k+1}$ are the Lagrange relaxation multipliers at the \textit{$k^{th}$} and the \textit{$(k+1)^{th}$} iteration, respectively; and $h\in g^k$, i.e. a sub-gradient of the objective function given in (\ref{eq:relaxed LP modified}) at the $k^{th}$ iteration. Also, $\alpha^{k}$ denotes the size of the step taken in the direction of the sub-gradient at the $k^{th}$ iterative step, and is updated as $\alpha^k=0.01/\left(k+1\right)$. This approach satisfactorily parallelizes FPGA routing and gives better results over VPR \cite{Betz}, but there are many inequality constraints that are violated for some cases. Furthermore, the minimum channel width requirement needs to be improved further. 

In the formulated LP given by (\ref{eq:LP}), $w_{e}$ is constant $\forall e\in E$. Therefore, minimizing the objective function (that is, the total path cost of FPGA routing) automatically minimizes the channel width (i.e. $\sum_{i=1}^{N_{nets}}x_{e,i}$). We start with a constant value of $W$, and then solve the optimization problem given by (\ref{eq:relaxed LP modified}). This gives us the total path cost and the channel width \big(also the inequality constraints violation from (\ref{eq:LP}), i.e. $\max (0,\sum_{i=1}^{N_{nets}}x_{e,i}-W)$\big). Next, we reduce the value of W and again follow the above steps to obtain a better local minima both for the total path cost and the channel width. If we obtain a reduced channel width, then the inequality constraints violation may reduce, increase or remain same. Usually, it decreases. Therefore, the above two problems are interlinked\footnote{For further detail, please see Section 3 of \cite{Hau}.}.

In our proposed work, we overcome the deficiencies of the existing approach of FPGA routing discussed in \cite{Hau} by appropriately calculating the Lagrange relaxation multipliers and the corresponding step sizes. We implement three different variants of sub-gradient based methods, namely, the Projected sub-gradient method (as done in \cite{Hau}), the Primal-Dual sub-gradient method, and the Deflected sb-gradient method. The difference among these variants include the different expression for the iterative update of the Lagrange relaxation multipliers and the corresponding step sizes. 

\subsection{Our algorithm}\label{sec:Our Algorithm}

Next, we discuss our algorithm that better exploits the problem properties. We use the Primal-Dual sub-gradient method because the LP given in (\ref{eq:relaxed LP modified}) is the Lagrange dual problem of the LP given in (\ref{eq:LP}), and hence, this method is a natural fit here. That is, the Primal-Dual sub-gradient method is useful because of the way the Lagrange relaxation multipliers are updated. That is,
{\vspace{-3mm}
\begin{align}\label{eq:Lagrangian multiplier}
& \lambda_{e}^{k+1}=\lambda_{e}^{k}+\alpha^{k}\max\left(0,\sum_{i=1}^{N_{nets}}x_{e,i}-W\right),
\end{align}
where $\sum_{i=1}^{N}x_{e,i}-W$ is a sub-gradient of the objective function at the $k^{th}$ iteration (the partial derivative of the objective function in (\ref{eq:relaxed LP modified})), and $\lambda_{e}^{0}=0 \: \forall e\in E$ is the most general initial guess \cite{Fisher}. 

Let us now compare (\ref{eq:Lagrangian multiplier}) with the update given in the Projected sub-gradient method (as discussed in the above paragraphs). For both the methods, if the inequality constraints are violated at the $k^{th}$ iteration, then the Lagrange relaxation multiplier at the $(k+1)^{th}$ iteration is incremented by $\alpha^k$ times the sub-gradient of the objective function at the $k^{th}$ iteration\footnote{For the Primal-Dual sub-gradient method this is obvious. For the Projected sub-gradient method, $\lambda^{k}$, $\alpha^{k}$, and $h$ all would be positive.}. Otherwise, for the Primal-Dual sub-gradient method, the value of the Lagrange relaxation multiplier at the $(k+1)^{th}$ iteration is the same as the $k^{th}$ iteration, while for the Projected sub-gradient method, it may change. In general, this works better because, if there is no constraints violation at the $k^{th}$ iteration, then the Lagrange relaxation multiplier at the $(k+1)^{th}$ iteration should remain the same.

Next, we discuss the choice of the step size. If the step size is too small, then the algorithm would get stuck at the current point, and if it is too large, the algorithm may oscillate between any two non-optimal solutions. Hence, it is very important to select the step size appropriately. The choice of step size can be either constant in all the iterations or can be reduced in each successive iteration. In our proposed scheme, the computation of step size involves a combination of the iteration number as well as the norm of the \textit{KKT} operator (Karush-Kuhn-Tucker operator) of the objective function at that particular iteration \cite{Guta} (instead of using the iteration number only, as given in \cite{Hau}). This ensures that the problem characteristic is used in the computation of the step size. That is,
\begin{align*}\label{eq:step size}
& \alpha ^{k}=\left ( 1/k \right )/\left \| T^{k} \right \|_{2},
\end{align*}
where \textit{k} is the iteration number, $T^k$ is the \textit{KKT} operator for the objective function of (\ref{eq:relaxed LP modified}), and $\left \| T^{k} \right \|_{2}$ is the 2-norm of $T^k$. 

The sub-gradient based methods are iterative algorithms, and hence, we need to check when to stop. There is no ideal stopping criterion for sub-gradient based methods, however, some possible measures that can be used \cite{Duchi} are discussed below (including our choice).
\begin{itemize}[noitemsep,topsep=0pt,parsep=0pt,partopsep=0pt]
	\item If at an iteration \textit{k}, $g^k\leq0$ and $\lambda^kg^k=0$, then we obtain the optimal point. Therefore, we stop here. However, this stopping criteria is achieved only if strong duality holds but, in case of our problem, there is weak duality\footnote{Detail of strong and weak duality can be found in \cite{Guta}.}. 
	
	\item Let at iteration \textit{k}, $f^*$ and $f^{k}_{best}$ are the optimal value and the best possible value, respectively, of the objective function, then the sub-gradient iterations can be stopped when $|f^{k}_{best}-f^{*}|\leq \epsilon$ (where $\epsilon$ is a very small positive number). In this criteria, the optimal value of the objective function is required in advance, which we do not have.
	
	\item In diminishing step size, as discussed at the start of Section \ref{Sec:Implementation}, when the step size becomes too small, the sub-gradient method would get stuck at the current iteration, and hence, we can stop sub-gradient iterations. However, for our problem, there is no proper criteria for deciding the lower limit of the step size.
\end{itemize}
As any of the above stopping criteria do not suit us, we stop our algorithm after a sufficient and fixed number of iterations, as used in \cite{Hau}.
\section{Experimental results}\label{Sec:Experimental Results}

Using the earlier algorithm, we perform experiments on a machine with single Intel(R) Xeon(R) CPU E5-1620 v3 running at 3.5 GHz and 64 GB of RAM. The operating system is Ubuntu 14.04 LTS, and the kernel version is 3.13.0-100. Our code is written in C++11 and compiled using GCC version 4.8.4, and the resulting compiled code is run using different number of threads. We compare our method with VPR \cite{Betz} and ParaLaR \cite{Hau}. For comparison purpose, VPR 7.0 from the Verilog-to-Routing (VTR) package and ParaLaR are compiled using the same GCC version.  Some parameters in the input-output pad and the configuration logic blocks (CLBs) of VPR and ParaLaR are modified to run them identical to our model. We test on MCNC benchmark circuits \cite{Yang}, which range from small sized to large sized logic blocks. We use an upper limit of $50$ for the number of iterations of the sub-gradient method.

In Table \ref{table1}, we compare the total path cost, the channel width, and the critical path delay (in nanoseconds) as obtained by our algorithm with VPR and ParaLaR. For sake of easy comparison, we call the total path cost as the total wire length here. These metrics are independent of the number of threads used, therefore, here we do not present the results for different number of threads (which is discussed in the following paragraph). If we look at the total wire length, then our algorithm gives average savings of 20.13\% over VPR, which is the same as for ParaLaR. If we look at the channel width, then our algorithm gives 22.12\% improvement over ParaLaR, which is the same as for VPR. Recall, the constraints violation is the difference of the channel width and the input W. Hence, this improves proportionally to the channel width improvement. Hence, our algorithm achieves minimum value for all the three parameters. 

Unfortunately, minimizing the channel width and total wire length causes our algorithm to incur some extra cost (very less) in terms of critical path delay.  We can see from Table \ref{table1} that this delay for our algorithm is on an average only $1.95\%$ higher than that of VPR, and on an average $3.87\%$ higher than that of ParaLaR. Hence, our algorithm incurs very little cost in terms of time.

\begin{table*}[!t]
	\def\arraystretch{1.2}
	\centering
	\caption{Comparison of quality of results between our algorithm, VPR, and ParaLaR.}\label{table1}\vspace{-2.5mm}
	\smallskip
	\begin{tabular}{|l|ccc|ccc|ccc|}
		\hline
		Benchmark     & \multicolumn{3}{c|}{Total wire length} & \multicolumn{3}{c|}{Channel width} & \multicolumn{3}{c|}{Critical path delay (ns)} \\
		& Proposed & VPR \cite{Betz} & ParaLaR \cite{Hau}  & Proposed & VPR \cite{Betz} & ParaLaR \cite{Hau} & Proposed & VPR \cite{Betz} & ParaLaR \cite{Hau}\\
		\hline
		Alu4 & 5087 & 6538 & 5087 & 35 & 38 & 48 &  7.01 & 7.43 &	6.89 \\ 
		Apex2 & 7927 & 10233 & 7928 & 48 & 50 & 59  & 7.90 &	8.23 & 7.50\\ 
		Apex4 & 5652 & 7190 & 5650 & 49 & 48 & 61  & 7.50 & 7.25 & 6.38\\
		Bigkey & 4173 & 4711 & 4173 & 18 & 26 & 23 & 3.69 & 2.67 &	4.30\\ 
		Clma & 50310 & 62086 & 50328 & 81 & 76 & 104  & 15.90 & 15.13 & 14.95\\ 
		Des & 7050 & 8892 & 7047 & 30 & 30 & 41  & 5.63 & 5.83 & 5.32 \\ 
		Diffeq & 4522 & 6299 & 4522 & 42 & 34 & 50  & 5.60 & 5.71 & 5.60\\
		Dsip & 4935 & 5952 & 4935 & 31 & 28 & 34  & 3.86 & 3.12 & 2.89\\ 
		Elliptic & 15198 & 19150 & 15202 & 58 &	54	& 79  & 9.30 & 9.02 & 9.25\\ 
		Ex1010 & 23277 & 29474 & 23268 & 67 & 62 & 81  & 11.90 & 10.83 & 12.31\\
		Ex5p & 4921	& 6289 & 4921 & 50 & 50	& 66  & 6.40	& 7.90	& 7.15\\
		Frisc & 19668 & 24095 & 19659 &	71 & 62 & 89  & 12.54 & 11.67 & 11.22\\
		Misex & 5229 & 6789 & 5230 & 47 & 44 & 57  & 6.70	& 6.01 & 6.42\\
		Pdc	& 30685 & 36803 & 30667 & 75 & 74 &	84  & 12.40 &	13.60 & 11.90\\
		S298 & 5208 & 6610 & 5208 & 37 &	40	& 49  & 12.90 &	10.24 & 11.54\\
		S38417 & 21705 & 28671 & 21707 & 50	& 48  &	86  & 8.05 &	8.50 & 8.47\\
		Seq & 7672 & 9691 & 7671 & 51 & 50	& 63  & 6.98 &	6.38 & 6.67\\
		Spla & 20404 & 25115 & 20402 & 66 &	64	& 86  & 11.34 & 12.50 & 10.61\\
		Tseng & 2436 & 3504 & 2436 &	34	& 38 &	47  & 5.27 & 5.70 & 5.27\\
		\hline
		\textbf{Total} & \textbf{246059} & \textbf{308092} & \textbf{246041} & \textbf{940} & \textbf{916} & \textbf{1207} & \textbf{160.87} & \textbf{157.72} & \textbf{154.64} \\ \hline 	
		\textbf{Average} & \textbf{12950.47} & \textbf{16215.37} & \textbf{12949.53} & \textbf{49.47} & \textbf{48.21} & \textbf{63.53} & \textbf{8.47} & \textbf{8.30} & \textbf{8.14}\\ \hline
	\end{tabular}
\end{table*}

Recall, the underlying goal of \cite{Hau} and us (we improve the algorithm in \cite{Hau}) is to efficiently parallelize the routing process. Hence, next we report results when using different number of threads in Table \ref{table2}. The benchmark dataset used is the same as discussed in the earlier paragraph (first column in Table \ref{table2}). Columns second through sixth give the absolute runtime and the remaining columns give the relative speedups. The speedup of execution with \textit{n} threads is calculated as
\begin{align*}
Speedup=\frac{Execution\ time\ with\ n\ threads}{Execution \ time\ with\ 1\ thread}.
\end{align*}

The symbols \textit {1X, 2X, 3X, 4X} in Table \ref{table2} refers to the execution of our algorithm with $1, 2, 3$ and $4$ threads, respectively. We can observe from this table that our algorithm (when we run it using $1$ thread) is $2.16$ times faster than that of VPR. It can also be observed from this table that when we use our algorithm with $2$ threads, on an average, speedups of upto $1.78$ times are obtained over the single thread execution. Similarly, using $3$ threads, on an average, speedups of upto $2.29$ times, and when using $4$ threads, on an average, speedups of upto $2.95$ times are observed.
	
We also calculate the speedup of ParaLaR, and compare it with our proposed method. We achieve almost similar speedups, and hence, we do not report this data here. Thus, our proposed method achieves similar parallelization as compared to ParaLaR.

\begin{table*}[!t]
	\def\arraystretch{1.2}
	\centering
	\caption{Execution time (in second) of VPR and our algorithm, and speedup when using different number of threads.}\label{table2}
	\smallskip
	\begin{tabular}{|l|ccccc|cccc|}
		\hline
		Benchmark     & \multicolumn{5}{c|}{Execution time (s)} & \multicolumn{4}{c|}{Speedup} \\
		& VPR & 1X  & 2X & 3X & 4X & 1X vs VPR &  2X vs 1X & 3X vs 1X & 4X vs 1X \\
		\hline
		Alu4 & 8.28 & 10.45 & 6.25 & 5.11 & 3.35 & 0.79 & 1.67 & 2.05 &	3.12 \\ 
		Apex2 & 8.58 & 32.49 & 17.15 & 11.53 & 8.52 & 0.26  & 1.89 & 2.82 & 3.81 \\ 
		Apex4 & 4.7 & 8.06 & 4.23 & 2.9 & 2.67 & 0.58  & 1.91 & 2.78 & 3.02 \\
		Bigkey & 2.81 & 0.97 & 0.67 & 0.7 & 0.61 & 2.90 & 1.45 & 1.39 &	1.59 \\ 
		Clma & 395.86 & 83.97 & 44.38 & 30.24 & 28.18 & 4.71 & 1.89 & 2.78 & 2.98 \\ 
		Des & 9.57 & 2.68 & 1.59 & 1.31 & 0.96 & 3.57  & 1.69 & 2.05 & 2.79 \\ 
		Diffeq & 6.54 & 2.43 & 1.43 & 1.27 & 0.87 & 2.69 & 1.7 & 1.91 & 2.79 \\
		Dsip & 4.72 & 0.76 & 0.61 & 0.52 & 0.49 & 6.21  & 1.25 & 1.46 & 1.55 \\ 
		Elliptic & 52.14 & 27.56 & 14.23 & 9.85 & 8.99 & 1.89  & 1.94 & 2.80 & 3.07 \\ 
		Ex1010 & 37.05 & 26.4 & 13.88 & 9.5 & 7.56 & 1.40 & 1.90 & 2.78 & 3.49 \\
		Ex5p & 6.38	& 4.42 & 2.49 & 1.74 & 1.6	& 1.44 & 1.78 & 2.54	& 2.76 \\
		Frisc & 56.86 & 9.75 & 5.53 & 3.88 & 3.33 & 5.83 & 1.76 & 2.51 & 2.93 \\
		Misex & 5.55 & 18.74 & 9.74 & 8.95 & 5.59 & 0.30  & 1.92 & 2.09 & 3.35 \\
		Pdc	& 306.69 & 107.2 & 56.41 & 51.27 & 27.75 &	2.86  & 1.90 &	2.09 & 3.86 \\
		S298 & 8.34 & 6.72 & 3.61 & 3.38 &	2.51 & 1.24  & 1.86 & 1.99 & 2.68 \\
		S38417 & 26.13 & 10.76 & 6.04 & 4.28 & 3.86  &	2.43  & 1.78 &	2.51 & 2.79\\
		Seq & 10.67 & 21.25 & 11.19 & 7.5 & 5.8	& 0.5  & 1.90 &	2.83 & 3.66\\
		Spla & 54.79 & 156.51 & 79.81 & 65.84 &	49.79 & 0.35  & 1.96 & 2.38 & 3.13\\
		Tseng & 1.74 & 1.53 & 0.93 & 0.83 & 0.56 &	1.14  & 1.65 & 1.84 & 2.73 \\
		\hline
		\textbf{Total} & \textbf{1007.4} & \textbf{532.65} & \textbf{280.17} & \textbf{220.6} & \textbf{163.17} & & & & \\ \hline 	
		\textbf{Average} & \textbf{53.02} & \textbf{28.03} & \textbf{14.75} & \textbf{11.61} & \textbf{8.59} & \textbf{2.16} & \textbf{1.78} & \textbf{2.29} & \textbf{2.95}\\ \hline
	\end{tabular}
\end{table*}

\section{Conclusions and future work}\label{Sec:Conclusion}

In this work, we extend the work of \cite{Hau} in proposing a more effective parallelized FPGA router. We use the LP framework of \cite{Hau} and use the Primal-Dual sub-gradient method with better update of the Lagrange relaxation multipliers and the corresponding step sizes. 

Experiments on the standard benchmarks show that using our algorithm gives improvements of upto 22\% in the standard metric of the minimum channel width as compared to ParaLaR \cite{Hau}, our parent algorithm (which is the same as in VPR \cite{Betz}, another standard algorithm). This proportionally reduces the constraints violation, which was a problem in ParaLaR.  We achieve the same total wire length (another standard metric) as ParaLaR. This is 20\% better than the corresponding data for VPR. Hence, our algorithm achieves minimum value for all the three parameters. Our algorithm incurs very less extra timing cost in terms of the critical path delay, and executing it in parallel gives speedups of upto $3$ times with $4$ threads (over our serial implementation).

The Lagrange relaxation technique that we use, is not always guaranteed to satisfy the corresponding constraints (as observed in Sections \ref{Sec:Implementation} and \ref{Sec:Experimental Results}). Hence, one future direction is to develop a Lagrange heuristic \cite{Lagrangian heuristics 1, Lagrangian heuristics 2, Lagrangian heuristics 3} specific to our problem to avoid this behavior. Another future direction involves experimenting on the Titan benchmark \cite{Titan}.

\ifCLASSOPTIONcaptionsoff
  \newpage
\fi

\end{document}